\documentclass[preprint, 3p, twocolumn]{elsarticle}

\usepackage{booktabs}
\usepackage{dcolumn}% Align table columns on decimal point
\usepackage{float}
\usepackage[export]{adjustbox} % allows left and right align of figures
\usepackage{wrapfig} % allows text to wrap around a figure
\usepackage{bm}% bold math
 \usepackage{booktabs}
 \usepackage{multirow}
\usepackage[dvipsnames]{xcolor}

\usepackage{graphicx}
\usepackage{booktabs}
\usepackage[normalem]{ulem}
 \useunder{\uline}{\ul}{}
 \usepackage{rotating}
 \usepackage{dsfont}
 \usepackage{textcomp}
\usepackage{gensymb}
\usepackage{color}
\newcommand{\bit}[1]{\textit{\textbf{#1}}}
\usepackage[switch]{lineno}
\usepackage{graphicx,xcolor}
\usepackage{epstopdf}
\usepackage{setspace}
\usepackage[T1]{fontenc}
\usepackage{amssymb,amsfonts,amsmath}
\usepackage{setspace}
\usepackage{subcaption}
\usepackage{mathtools}
\usepackage{etoolbox}
\biboptions{sort&compress}
\newcolumntype{Y}{>{\centering\arraybackslash}X}
\begin{document}

%\preprint{APS/123-QED}
%\linenumbers
\title{Dewetting Characteristics of Contact Lenses Coated with Wetting Agents}% Force line breaks with \\
\author[vinny,harvard,wyss,equal]{V. Chandran Suja}
\ead{vinny@seas.harvard.edu}
% \affiliation{%
%  Department of Chemical Engineering, Stanford University, CA - 94305, USA
% }
% \affiliation{%
%  School of Engineering and Applied Sciences, Harvard University, MA - 01234, USA
% }
% \affiliation{%
% Wyss Institute for Biologically Inspired Engineering, 52 Oxford St, Cambridge, MA 02138, USA
% }
% \affiliation{%
%  Equal contribution
% }
\author[vinny,equal]{A. Verma}
% \affiliation{%
%  Department of Chemical Engineering, Stanford University, CA - 94305, USA
% }
% \affiliation{%
%  Equal contribution
% }
\author[vinny]{E. J. L. Mossige}
\author[vinny]{K. W. Cui}
\author[vinny]{V. Xia}
% \affiliation{%
%  Department of Chemical Engineering, Stanford University, CA - 94305, USA
% }
\author[jnj]{Y. Zhang}
\author[jnj]{D. Sinha}
\author[jnj]{S. Joslin}
% \affiliation{%
%  Johnson \& Johnson Vision Care Inc., Jacksonville, FL 32256, USA
% }
\author[vinny]{G.G. Fuller}
\ead{ggf@stanford.edu}
% \affiliation{%
%  Department of Chemical Engineering, Stanford University, CA - 94305, USA
% }

\address[vinny]{Department of Chemical Engineering, Stanford University, CA - 94305, USA}
\address[harvard]{School of Engineering and Applied Sciences, Harvard University, MA - 01234, USA}
\address[wyss]{Wyss Institute for Biologically Inspired Engineering, 52 Oxford St, Cambridge, MA 02138, USA}
\address[equal]{Authors contributed equally}
\address[jnj]{Johnson \& Johnson Vision Care Inc., Jacksonville, FL 32256, USA}

%\date{}% It is always \today, today,

\begin{abstract}
\textit{Hypothesis}

 Although wetting agents have been developed to limit tear film dewetting over contact lenses, systematic analyses correlating wetting agents properties to mechanisms of the tear film destabilization are not readily available. Clarifying destabilization characteristics across key physio-chemical variables will provide a rational basis for identifying optimal wetting agents. \\

\textit{Experiments}

We employ an in-house, \emph{in vitro} platform to comprehensively evaluate drainage and dewetting dynamics of five wetting agents across seventeen different formulations and two model tear film solutions: phosphate-buffered saline (PBS) and artificial tear solution (ATS). We consider the film thickness evolution, film thickness at breakup, dewetted front propagation, and develop correlations to contact angle to compare the samples. \\

\textit{Findings}

Zwitterionic wetting agents effectively stabilize the tear film by reducing the film thickness at the onset of dewetting, and delaying dewetted region propagation across the lens. Furthermore, tuning wetting agent surface concentrations in binary mixtures can enhance wetting characteristics. Finally, despite disparities in wetting agent molecular properties, the time to dewet $50\%$ of the lens scales linearly with the product of the receding contact angle and contact angle hysteresis. Hence, we fundamentally establish the importance of minimizing both the absolute contact angle values and contact angle hysteresis for effective wetting performance. 

\end{abstract}

\maketitle
% \section{To Do: Winter 2021}
% \begin{itemize}
%     \item Analyze the drainage and dewetting data from the experiments (Make discoveries/correlations)
%     \item Literature review
%     \item for Archana: Read up more on dewetting physics for more clear justifications
% \end{itemize}

\section{Introduction} 
The tear film is a colorless liquid coating that maintains ocular health by lubricating the eye and simultaneously offering protection from allergens, debris, and pathogens. Typically, a class of glycoproteins called mucins lubricate the ocular surface through the retention of water; this creates a hydrophilic surface that promotes tear film stability \cite{rabiah2020understanding, mantelli2008functions}. However, this surface (and thus the associated tear film stability) can be disrupted upon insertion of contact lenses due to differences in interfacial properties \cite{lu2014tear}. 

Currently, silicone hydrogels (SiHy) are the preferred material for contact lenses, primarily due to their increased oxygen permeability\cite{nandu1993process}. However, this adapted polymer is significantly more hydrophobic than traditional hydrogels \cite{havuz2020vitro, read2011dynamic}, raising concerns of an increased propensity for lipid deposition and reduced surface wettability \cite{tighe2000silicone, pucker2010vitro}. Both adverse surface wettability and lipid absorption to lens materials are known to increase the probability of tear film break up and promote dewetting \cite{bhamla2015dewetting, bhamla2016instability}. 

Dewetting refers to the spontaneous withdrawal (negative spreading coefficient) of a liquid film from a surface due to the thermodynamic driving force for a system to exist in the lowest total free energy state\cite{de2013capillarity}. A dewetting-prone tear film is concerning since dewetting results in a number of ophthalmic disorders including dry eyes, poor vision, and discomfort. Contact lens discomfort and dryness symptoms are known consequences of tear film instability, affects around 50\% of all contact lens users\cite{nichols2005self}, and is the primary reason for patient discontinuation of contact lens use\cite{nichols2006tear}. Further, the dewetting of the pre-contact lens tear film can lead to an uncomfortable and  painful experiences when removing contact lens \cite{de2013capillarity}. These pressing issues necessitate innovation targeting stabilization of tear films when interacting with contact lens surfaces. 

Contact lens manufacturers have investigated a number of routes to improve the wettability and comfort of SiHy lenses including adjustments in lens materials, addition of surface active agents to lens packaging solutions, lens surface modifications, and incorporation of wetting agents \cite{nichols2006tear}. The scope of this study focuses on an adaption of the last approach due to the tunability and ease of manufacturing of wetting agent coatings. Despite the availability of a several commercial lenses with wetting agents, there lacks a thorough and systematic evaluation that links the macroscopic drainage and dewetting dynamics with the microscopic properties of these different additives. A successful wetting agent prolongs tear film stability and leads to low critical thicknesses, defined as the thickness of the film at the onset of the first dewetting event \cite{SHARMA1990433}. Clinically, the “contact lens surface wettability” (CLW) metric is used to evaluate the combined tear-film spreading and surface dewetting characteristics \cite{varikooty2012estimating}. A variety of related grading scales and methods have been defined to measure CLW \cite{woods2011development, stern2004comparison, long2000six, morgan2002comparative, zhang2011novel, varikooty2012estimating}, but measurements of \textit{in vivo} CLW typically consider precorneal lipid layer interference patterns, dark reflection spots, tear film breakup, and re-establishment of the tear film following a blink \cite{varikooty2012estimating, terry1993cclru}. Existing \textit{in vivo} methodologies used to evaluate contact lens wettability are valuable clinical tools; however, due to person-to-person variations in tear film characteristics and the multifaceted nature of \textit{in vivo} tear film instability, \textit{in vivo} measurements are not suitable for systematically evaluating mechanisms of wetting agent performance. %tear film dynamics and stability a complex challenge \cite{varikooty2012estimating}.
% Check varikooty2012estimating

%Contact lens coating formulations result from a complex optimization of wettability, biocompatibility, oxygen permeability, water content\cite{fonn2007targeting}, ease-of-use, and antifouling\cite{musgrave2019contact}.

Beyond identifying specific wetting agents that lead to the ideal drainage and dewetting properties, we seek to more generally identify physical design parameters that can help rationalize the performance of wetting agents. One such parameter is the contact angle. Contact angles represent the angle where a liquid-vapor interface meets a solid surface, and quantitatively describe the wettability of a solid surface via the Young equation \cite{de2013capillarity}. A greater contact angle corresponds to lower wettability (a fully wettable surface has a contact angle of 0°). Using the contact angle to quantify surface wettability is a common occurrence in dewetting physics literature at large; however, comparatively fewer studies have systematically applied this metric to tear films over contact lenses \cite{meadows2004dynamic, fagehi2013contact, campbell2013applicability, read2011dynamic}. We additionally seek to correlate the molecular properties of a wetting agent to the ability to prevent dewetting and understand how wetting agent concentration impacts the drainage and dewetting profiles. 

To accomplish the above goals, we performed \textit{in vitro} dewetting experiments using an in-house built platform for monitoring tear film drainage and dewetting: the interfacial dewetting and drainage optical platform (i-DDrOP) \cite{bhamla2016ddrop}. We used two model tear film solutions for the dewetting experiments -  phosphate buffered saline (PBS) and artificial tear film solution (ATS). ATS contains two vital tear film proteins, lysozyme and mucin; it is commonly used to mimic the \textit{in vivo} tear film for contact lens dewetting experiments   \cite{bhamla2016instability,rabiah2019influence}. By contrast, PBS lacks organic content, which aids in evaluating characteristics of wetting agents independent of the extraneous interactions from organic tear film components. Our experiments utilize the same base lens material, enabling us to isolate the effect of the surface coatings from the bulk material properties, such as contact lens permeability and cross link density \cite{liu2020tuning}. Finally, through this work, we establish a reproducible methodology and simple metric to identify suitable wetting agents for contact lenses.
 
The remainder of the manuscript is organized as follows. In section \ref{sec:Materials}, we introduce the array of surface coatings (\ref{subsec: MatContactLenses}), along with the experimental setup (\ref{iddrop}) and analyses (\ref{filmThickness}). Subsequently, in section \ref{Results} we discuss four primary areas of focus to characterize the influence of wetting agents on the tear film dewetting dynamics: film thickness at the inception of dewetting (\ref{filmThickness}), the dynamics of the dewetted front (\ref{dewetting}), concentration-dependent trends (\ref{WAconc}), and correlations to the surface energy and contact angle (\ref{surfEnergy}). Finally, in section \ref{conc}, we summarize our findings and suggest steps for further research.

% \begin{itemize}
%     \item Suggested papers: \cite{fatt1991observations,bhamla2016instability,rabiah2019influence,king2018mechanisms,doane1989instrument, braun2012dynamics, nichols2006tear, havuz2020vitro, SHARMA1990433}
% \end{itemize}

\begin{table*}
\caption{Tested wetting agents and their physio-chemical characteristics including the wetting agent type, material type/function, surface concentration, contact angle (CA) hysteresis, advancing dynamic contact angle (Adv. DCA), and equilibrium contact angle (Eq. CA).} \label{tab:Materials} %The surface concentration designations are assigned as follows\textcolor{red}{: low - x, medium - y, and high - z, CROSS CHECK THE CONCENTRATION OF GMA WITH JNJ, Recheck the eq. contact angle measurements - usually eq. contact angle is in between the adv. and rec. contact angle. Are the measured negative contact hystereses real?}. The contact angle data is the average of 5 independent measurements.
\begin{tabular}{@{}lcccccc@{}}
\toprule
No. & Wetting Agent                                                                                                         & Material/Function                                                                   & Conc.  & \begin{tabular}[c]{@{}c@{}}CA Hysteresis \\ ($\theta_a-\theta_r$) \end{tabular}    & \begin{tabular}[c]{@{}c@{}}Adv.\ DCA \\ ($\theta_a$) \end{tabular}  & \begin{tabular}[c]{@{}c@{}}Eq.\ CA \\ ($\theta_e$) \end{tabular}   \\ \midrule
1   & \multirow{3}{*}{\begin{tabular}[c]{@{}c@{}}Ampholyte-1\end{tabular}}                               & \multirow{3}{*}{\begin{tabular}[c]{@{}c@{}}Amino Acid,\\ Zwitterionic\end{tabular}} & Low    & 10.0          & 34       & 35.3   \\  \cmidrule(l){4-7} 
2   &                                                                                                                       &                                                                                     & Medium & 0.7           & 39.7     & 24.5   \\  \cmidrule(l){4-7} 
3   &                                                                                                                       &                                                                                     & High   & 0.3           & 29.3     & 22.0   \\ \midrule
4   & \multirow{3}{*}{\begin{tabular}[c]{@{}c@{}}Zwitterion-1\end{tabular}}            & \multirow{3}{*}{\begin{tabular}[c]{@{}c@{}}Phospholipid,\\ Zwitterionic\end{tabular}}                                                       & Low    & 3.3           & 49.8     & 36     \\  \cmidrule(l){4-7} 
5   &                                                                                                                       &                                                                                     & Medium & 3.4           & 42.8     & 37.8   \\  \cmidrule(l){4-7} 
6   &                                                                                                                       &                                                                                     & High   & 4.4           & 53       & 14.3   \\ \midrule
7   & \multirow{3}{*}{\begin{tabular}[c]{@{}c@{}}Ampholyte-2\end{tabular}}                              & \multirow{3}{*}{\begin{tabular}[c]{@{}c@{}}Amino Acid,\\ Zwitterionic\end{tabular}} & Low    & 10.7          & 43       & 35.5   \\  \cmidrule(l){4-7} 
8   &                                                                                                                       &                                                                                     & Medium & 8.0           & 39       & 30.4   \\  \cmidrule(l){4-7} 
9   &                                                                                                                       &                                                                                     & High   & 8.3           & 51       & 30.3   \\ \midrule
10  & \multirow{3}{*}{\begin{tabular}[c]{@{}c@{}}PEO\end{tabular}} & \multirow{3}{*}{\begin{tabular}[c]{@{}c@{}}Antifouling,\\ Nonionic\end{tabular}}                                                        & Low    & 21.8          & 58.4     & 47.4   \\  \cmidrule(l){4-7} 
11  &                                                                                                                       &                                                                                     & Medium & 22.0          & 66.2     & 54.6   \\  \cmidrule(l){4-7} 
12  &                                                                                                                       &                                                                                     & High   & 19.0          & 61       & 49.2   \\ \midrule
13  & Glycerol Monomethacrylate (GMA)                                                                                       & \begin{tabular}[c]{@{}c@{}}Hydrophylic, \\ Nonionic \end{tabular}                                                                         & Medium & 12.6          & 62.1     & --     \\ \midrule
14  & \multirow{2}{*}{Ampholyte-1/GMA}                                                                                           & {\multirow{2}{*}{Mixture}}                                        & 50/50  & 3.2           & 39.3     & 39.0   \\  \cmidrule(l){4-7} 
15  &                                                                                                                       & \multicolumn{1}{l}{}                                                                & 75/25  & -0.6          & 34.1     & 32.7   \\ \midrule
16  & \multirow{2}{*}{Zwitterion-1/GMA}                                                                                             & \multirow{2}{*}{Mixture}                                                            & 50/50  & 0.9           & 45.3     & 26.0   \\  \cmidrule(l){4-7} 
17  &                                                                                                                       &                                                                                     & 75/25  & -1.9          & 38.3     & 30.3   \\ \midrule
18  & None                                                                                                                  & \begin{tabular}[c]{@{}c@{}}SiHy lens - 1 \\ (Material Control)\end{tabular}                                                               & --     & 22.8          & 67.4     & 56.6   \\ \midrule
19  & None                                                                                                                  & \begin{tabular}[c]{@{}c@{}}SiHy lens - 2 \\ (In-use Control)\end{tabular}            & --     & 20.7          & 62.7     & 59.9   \\ \bottomrule
\end{tabular}
\end{table*}

\begin{figure*}[!th]
\includegraphics[width=\linewidth]{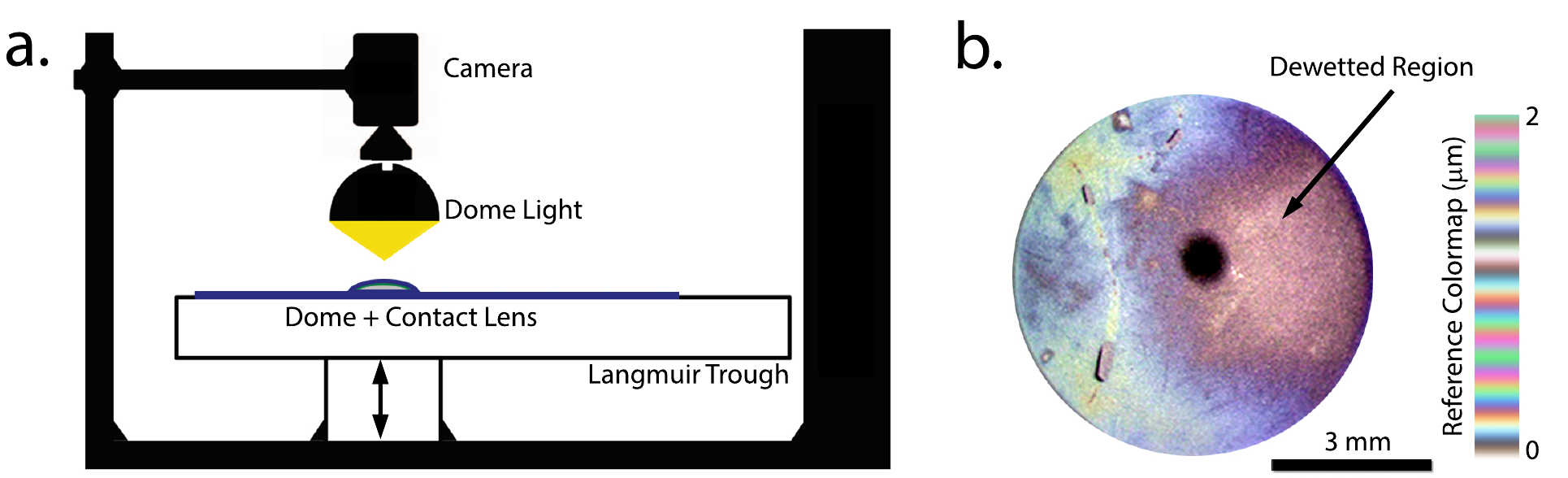} 
\caption{Schematic of the Interfacial Dewetting and Drainage Optical Platform (i-DDrOP) used for the \textit{in vitro} experiments  ({\bf a.}) The i-DDrOP setup with the labeled components. Image adapted from Bhamla \textit{et al.} \cite{bhamla2016instability} ({\bf b.}) \textit{In vitro} interferograms obtained over contact lenses. See \bit{Supplementary Video V1} for a typical video obtained through the camera.}
\label{fig:InvivoTearFilm_ExperimentalSetup}
\end{figure*}

\section{Materials and Methods}\label{sec:Materials}
\subsection{Contact lenses and wetting agents}\label{subsec: MatContactLenses}
Contact lenses modified with five unique materials were obtained from Johnson \& Johnson. Using a novel photo-cure process, the modifying agents were incorporated into SiHy lens material to improve wettability (Table \ref{tab:Materials}). Lens properties such as water content and wettability were evaluated to determine the changes after modification. Thirteen of the tested lenses (samples 1-13) have varied surface concentrations of five, single-component wetting agents: Ampholyte-1, Ampholyte-2, Zwitterion-1, Polyethylene oxide (PEO), and Glycerol Monomethacrylate (GMA). The two ampholyte coatings are derived from distinctly different amino acid groups, which thus lead to their varied macroscopic properties. Zwitterion-1 is a phospholipid polymer. The two top performing wetting agents from this set (Ampholyte-1 and Zwitterion-1) were mixed with GMA at two different ratios (samples 14-17). The remaining two samples (18-19) serve as controls: the material control is the same lens as those in samples 1-17 with no additional wetting agent added and the in-use control is a commercially available lens.

Wettability of lenses was determined by using the dynamic contact angle (DCA) (Table \ref{tab:Materials}). DCA was determined by the Wilhelmy plate method, using a Cahn DCA-315 instrument at room temperature with deionized water as the probe solution. The experiment was performed by dipping the lens specimen of known parameter into the packing solution of known surface tension while measuring the force exerted on the sample due to wetting by a sensitive balance. The advancing contact angle of the packing solution on the lens is determined from the force data collected during sample dipping. The receding contact angle is likewise determined from force data while withdrawing the sample from the liquid. Contact angle hysteresis is defined as the difference between the advancing and receding contact angles and may be used as a qualitative measure of surface roughness and heterogeneity, although other factors may be involved. On the relative basis, a surface with a larger contact angle hysteresis would be expected to exhibit greater surface roughness and heterogeneity.

\subsection{Interfacial dewetting and drainage optical platform experiments (i-DDrOP)}\label{iddrop}
\textit{In vitro} tear film dewetting experiments were performed using the Interfacial Dewetting and Drainage Optical Platform (i-DDrOP) \cite{bhamla2016instability, bhamla2016ddrop} shown in Fig.\ref{fig:InvivoTearFilm_ExperimentalSetup}d. The i-DDrOP consists of a dome (mimicking the eyeball, radius of curvature $= 8.6\;mm$) for holding a contact lens, a trough for holding a solution mimicking the tear liquid, a diffuse dome light (CCS America) to uniformly illuminate and create thin film interference patterns on the lens, and a camera (Thor Labs DCU223C) to record the information.
%Endre: Radius of curvature of dome? Frame rate of camera used for image acquisition?

For every i-DDrOP experiment reported herein, a selected contact lens was placed on the dome and submerged in the tear-like solution. At the start of the experiment, the contact lens (on the dome) was placed $1\;mm$ below the air-solution interface. Subsequently, contact lens was moved $2\;mm$ towards the air-liquid interface by lowering the trough, trapping a thin liquid film over the contact lens, analogous to the tear film coating post-blinking. The dome and adhered lens was then raised to the liquid surface, allowing this thin film to drain as it would over the course of a blink cycle. Recording the interference patterns over the contact lens at 30 fps  (Fig.\ref{fig:InvivoTearFilm_ExperimentalSetup}e) enables monitoring of the tear film drainage. All experiments were uncovered (free evaporation) and at room temperature ($20\ ^oC$).

    \begin{figure*}[!th]
    \centering
    \includegraphics[width=\linewidth]{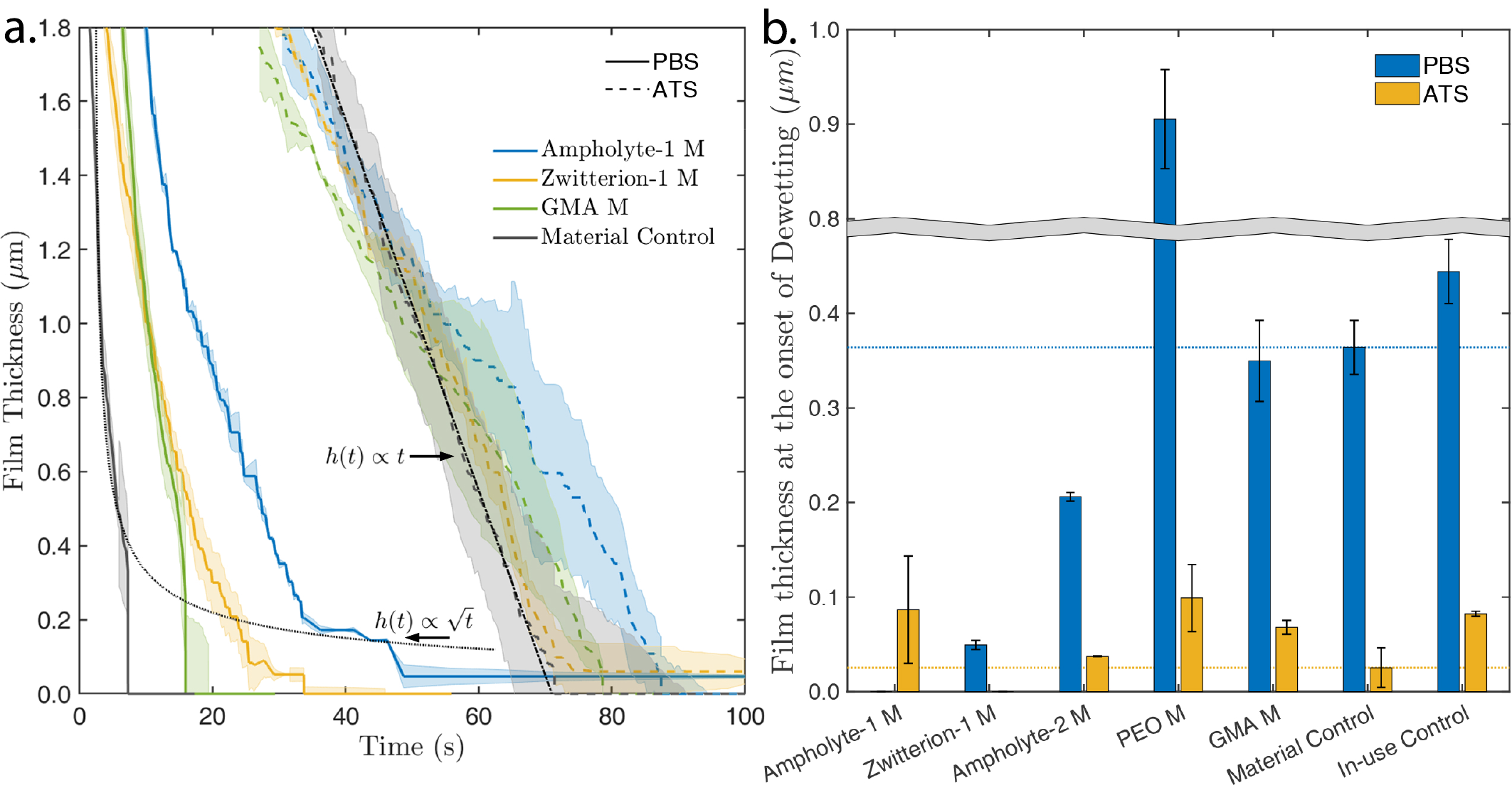}
    \caption{{\bf a.} Drainage profiles of select lenses. Shown here are Zwitterion-1 coated lenses, which has superior wetting characteristics as compared to the uncoated material control base lens in both PBS and ATS, and GMA coated lens which performs worse than the uncoated lens in both PBS and ATS. Films draining under the sole influence of capillary forces are expected to decay as $h(t) \propto t^{-1/2}$, whereas evaporation driven film drainage goes as $h(t) \propto t$. {\bf b.} Average thickness at the onset of dewetting for the tested wetting agents and control lenses. The top and bottom horizontal lines indicate the film thickness at dewetting for the material control lens in PBS and ATS respectively. The grey region indicates a break in the y-axis.}
    \label{fig:DrainageProfiles}
    \end{figure*}

\subsubsection{Contact lens preparation}\label{CLprep}
In order to leach out undesired materials present in the contact lens storage solution, we washed the contact lens following protocols established by Rabiah \textit{et al.} \cite{rabiah2019influence}. We implemented a four step wash protocol using 1x PBS in 24 well plates on a shaker set to intermediate, where the PBS solution was replaced four times (after 15 minutes, after another 30 minutes, followed by overnight soaking, and 30 minutes prior to experiments). 

\subsubsection{Model tear film solutions}\label{modelTear}
The film drainage over lenses were tested in two different solutions: phosphate-buffered saline (PBS), purchased from Gibco (Thermofisher Scientific), and artificial tear solution (ATS). 

PBS was chosen to decouple the effects of proteins and other substances in tear films \cite{VIDALROHR2018117} from the effects of the wetting agent. Having a simple buffer as the solvent allows for a more direct correlation of molecular properties of the wetting agent to its associated macroscale interfacial dewetting properties. 

To prepare the ATS, 7.5 mg bovine mucin (499643, EMD Millipore) was added to 50 mL of 1x PBS and mixed until completely dissolved. 100 mg of lysozyme (L6876,
Sigma Aldrich) was then added and stirred until completely dissolved. 

At $25\ ^oC$ temperature, dynamic oscillatory shear measurements using an interfacial stress rheometer at a strain of 1.5\% and frequency of 1 Hz revealed a elastic and loss surface shear modulus of $40$ and $10$ $mN/m$ for ATS \cite{rabiah2019influence}. PBS has no measurable interfacial rheology. ATS has a viscosity of 0.968 cP, while PBS has a viscosity of 0.933 cP \cite{rabiah2019influence}. 

\subsection{Film thickness evolution and dewetting analysis}\label{subsec:ThicknessReconstruction}
The spatiotemporal evolution of film thickness over the contact lenses was obtained by mapping the colors in the recorded interference patterns to physical thicknesses. This was achieved using the classical light intensity - film thickness relations \cite{sheludko1967thin, suja2020hyperspectral, suja2020single} assuming homogeneous and non-dispersive films (Fig.\ref{fig:InvivoTearFilm_ExperimentalSetup}c). A Python $2.7$ based software developed in-house \cite{frostad2016dynamic} aids in thickness mapping and visualizing the thickness profiles.

The dewetted area analysis utilizes in-house MATLAB codes that isolate the wetted from dewetted regions by thresholding and binarizing interference patterns from the experiments (Fig.\ref{fig:InvivoTearFilm_ExperimentalSetup}e). The ratio of the wetted area to the total analyzed area was tracked as function of time to obtain the dewetting profiles. See \bit{Supplementary Materials Fig. S1} for more details.

\section{Results and Discussion}\label{Results}  
\subsection{Film thickness evolution leading up to dewetting}\label{filmThickness}
Wetting agents have a minimal influence in the film drainage up until a few hundred nanometers in thickness. (Fig.\ref{fig:DrainageProfiles}a). At larger thicknesses, film drainage profiles are almost entirely dictated by the combined effect of tear film rheological properties, capillary forces and evaporation. Note that gravity has a relatively minor influence on film drainage, as the characteristic Bond number in our experiments $Bo = \frac{\rho g R h_0}{\gamma}\sim 0.02$, for density $\rho \sim 1000\;kg/m^3$, acceleration due to gravity $g \sim 10\;m/s^2$, dome radius of curvature $R \sim 0.01\;m$, reference initial film thickness $h_0 \sim 10\;\mu m$ and surface tension $\gamma \sim 0.05\; N/m$. Films draining under the sole influence of capillary forces evolve as $h(t) \sim t^{-1/2}$, whereas evaporation drives a linear decay in film thickness as $h(t) \sim t$. A simple analysis of the drainage scaling laws shows that the relative importance of these two processes vary during film drainage, with evaporation being the dominant film thinning mechanism beyond a critical time of $t_c \approx 30\ s$ (see \bit{Appendix} for details). As seen in Fig.\ref{fig:DrainageProfiles}a, the material control lens in PBS drains and dewets well before this critical time and exhibits a square-root drainage profile that is characteristic of capillary drainage. On the other hand, the material control in ATS continues to drain past this critical time, and manifests a linearly decaying film thickness profile that is characteristic of evaporation. 

%PBS ($Bq = \frac{\mu_s}{\mu h_0} \rightarrow 0$) has a free slip interface, which facilitates capillary driven drainage to the dewetting thickness before the critical time (beyond which evaporation becomes dominant). On the other hand,  ATS ($Bq \rightarrow \infty$) has a no slip interface, which retards capillary driven drainage. As a result, evaporation dominates film drainage. This manifests as linearly decaying film thickness in the ATS experiments.  

\begin{figure*}
    \includegraphics[width=\linewidth]{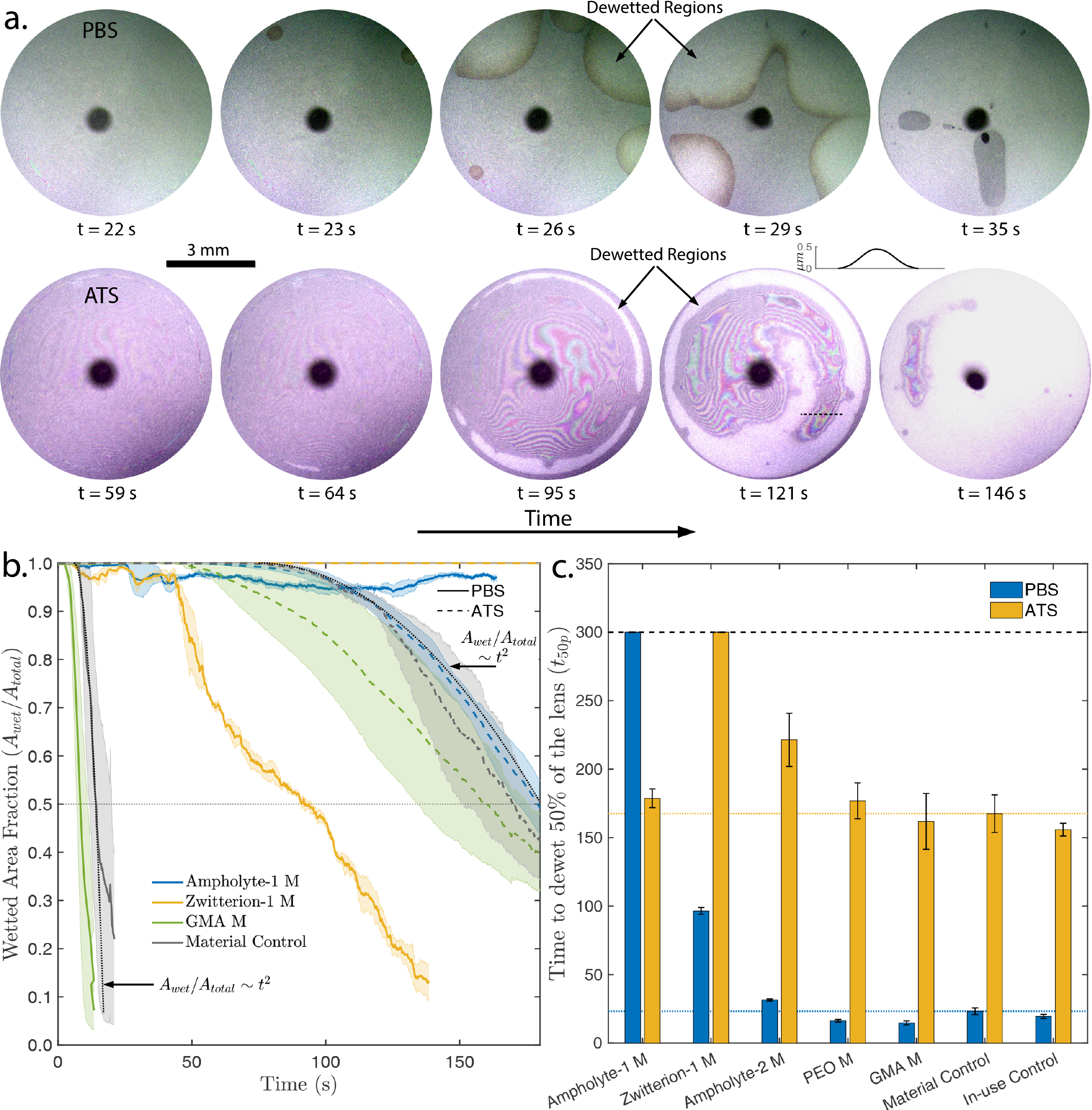}
    \caption{Dewetting dynamics of contact lenses with wetting agents. {\bf a.} Representative snapshots showing the dewetting dynamics of the material control lens in ATS and PBS. The inset profile shows the film thickness along the dotted line indicated in the snapshot with a time stamp of $t=121\;s$. {\bf b.} Dewetting profiles of selected lenses. Shown here are Zwitterion-1 coated lenses, which have superior wetting characteristics as compared to the uncoated material control base in both PBS and ATS, and the GMA coated lens, which performs worse than the uncoated lens in both PBS and ATS. The horizontal dashed line indicates a wetted fraction of $0.5$. {\bf c.} Time taken for 50\% of the contact lens surface area to dewet ($t_{50p}$). The middle and bottom dotted horizontal lines indicate $t_{50p}$ for the material control lens in PBS and ATS respectively. The top horizontal dashed line indicates the maximum duration of the dewetting experiments.} \label{fig:dewettingProfiles}
\end{figure*}    

Once the film decays to a critical thickness $hc$, it is prone to deweting. Theoretically, thick films are known be unstable below a critical thickness of $hc = 2\sqrt{\frac{\gamma}{\rho g}} \sin\left(\frac{\theta_e}{2} \right)$, where $\gamma$ is the surface tension and $\sqrt{\frac{\gamma}{\rho g}}$ is the capillary length \cite{redon1991dynamics}. In the absence of finite amplitude disturbances, such metastable films thin to mesoscopic ($\sim 1 \mu m$) thicknesses before spontaneously dewetting under conjoining forces exerted between the liquid-air and solid-liquid interfaces\cite{kheshgi1991dewetting,mitlin1993dewetting}. 

As the critical thickness is closely related to the physical characteristics of wetting agents and the tear film-air interface, we observe noticeable differences between the different systems (Fig.\ref{fig:DrainageProfiles}b). In PBS, the material control begins to dewet at a thickness of $\sim 350 nm$. The in-use and material controls coated with GMA dewet at a comparable thickness. On the other hand, addition of the antifouling agent PEO leads to an undesirable dramatic increase in the dewetting thickness. Zwitterionic wetting agents are observed to improve wetting behavior: addition of Ampholyte-2 resulted in a two-fold reduction and Zwitterion-1 in a seven-fold reduction in dewetting thickness. Interestingly, Ampholyte-1 is observed to resist dewetting over the time scale of an experiment. In ATS, most of the tested lenses have a comparable dewetting thickness as that of the control lens ($\sim 50 nm$). Notably, Zwitterion-1 coated lenses did not dewet in the time scale of the experiment. 

The capability to tune the critical dewetting thickness is one of the most important and useful facets of using wetting agents. An ideal contact lens wetting agent should lower the critical deweting thickness and prolong the dewetting time, thus ensuring the contact lens remains wetted completely between blink cycles. In this regard, we observe that Ampholyte-1 and Zwitterion-1 coated lenses perform well in both PBS and ATS.

\begin{figure*}
\includegraphics[width=\linewidth]{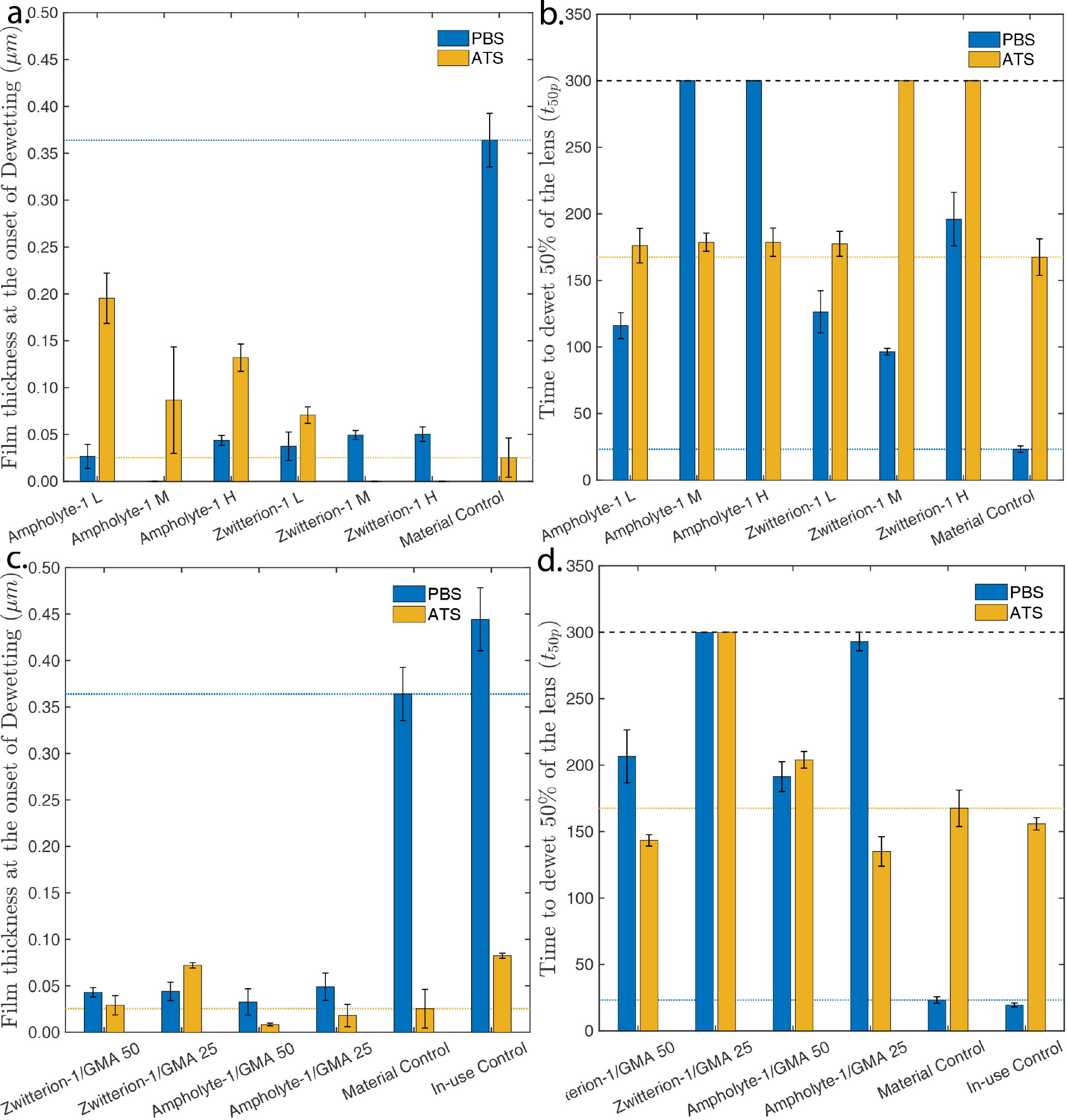}
\caption{Effect of wetting agent surface concentration and wetting agent mixtures. {\bf a.} Film thickness at dewetting for different wetting agent concentrations. {\bf b.} $t_{50p}$ for different wetting agent concentrations. {\bf c.} Film thickness at dewetting for different wetting agent mixtures {\bf d.} $t_{50p}$ for different wetting agent mixtures. The dotted horizontal lines indicate the measurements corresponding to the control the material control lens in ATS and PBS, while the dashed black horizontal line indicates the maximum duration of the dewetting experiments.} \label{fig:EffectOfConcMixtures} 
\end{figure*}

\subsection{Propagation dynamics of the dewetted front}\label{dewetting}
In many situations (e.g. following the deposition of lipids and proteins), the tear film can dewet during a blink cycle \cite{king2018mechanisms,lorentz2007lipid}. Wetting agents also play a crucial role post-dewetting by delaying the propagation of dewetted regions over the contact lens, thus maximizing the wetted lens surface area during a blink cycle. 

Representative snapshots from the material control lens experiments that depict qualitative dewetting characteristics observed across all lenses are shown in Fig.\ref{fig:dewettingProfiles}a. In both PBS and ATS, we observe the nucleation and growth dewetting mechanism \cite{de2013capillarity}. The growth rate of dewetted regions is highly dependent on the model tear film solution. For example, dewetted regions grow almost an order of magnitude slower in ATS than in PBS. The dewetted front propagates via circular arcs in PBS, indicating that capillary forces dominate the dewetting dynamics (see \ref{subsec:CorrelationSurfacEnergy} for more details). In the ATS trials, however, the dewetted fronts are irregularly shaped. Additionally, a rim that is characteristic of a dewetting front that accumulates liquid is absent in ATS, seen from the vibrant interference patterns that represent sub-micron film thickness at the dewetting front (e.g. Fig.\ref{fig:dewettingProfiles}a. ATS, $t=121\;s$). Thus, capillary forces play weaker role in the dynamics, likely due to the high interfacial viscoelasticity of ATS \cite{rabiah2019influence}. As is the case with film drainage (Fig. \ref{fig:DrainageProfiles}b), the uniform thinning of the film and the absence of a characteristic rim indicates evaporation dominates the dewetting dynamics in ATS.  

To quantify the influence of the wetting agents on the dewetting dynamics, we track the temporal evolution of the wetted area fraction of the lens ($A_{wet}/A_{total}$), shown in Fig.\ref{fig:dewettingProfiles}b (see \bit{Supplementary Video V1} for a dynamic visualization). $A_{wet}$ is the wetted area of the lens and $A_{total}$ is the maximum area of the lens that can dewet (i.e. area of the lens above the model tear solution in Langmuir trough). Initially, the wetted area fraction equals one for all lenses. Upon the appearance of a nucleated dewetted spot, $A_{wet}/A_{total}$ decays. At early times, $A_{wet}/A_{total} \sim t^{-1/2}$, indicating a linear growth for the effective radius of the dewetting front. The decay rate of the wetted area depends both on the model tear film solution and the wetting agent (Fig.\ref{fig:DrainageProfiles}b). At longer times, the decay rate decreases due to the finite size of the domain. Note that as mentioned in previous section, the lenses coated with Ampholyte-1 in PBS and Zwitterion-1 in ATS do not dewet on the time scale of the experiments.   

To practically compare the dewetting performance across surface coatings, we extract $t_{50p}$: the time at which the wetted area fraction equals $0.5$ (shown by the dotted line in Fig.\ref{fig:dewettingProfiles}b). From Fig.\ref{fig:dewettingProfiles}c, we observe the $t_{50p}$ of lenses in ATS are up to seven-fold longer as compared to PBS. The only exceptions are the lenses that do not dewet on the time scale of the experiments. Overall, zwitterionic Ampholyte-1 and Zwitterion-1 coated lenses effectively delay the propagation of dewetted regions across PBS and ATS. As compared to the control lens, Ampholyte-1 has a 10-fold higher $t_{50p}$ in PBS, and Zwitterion-1 has four-fold and a two-fold higher $t_{50p}$ in PBS and ATS respectively. Ampholyte-2 has a $t_{50p}$, which is about $50\%$ higher than the control lens in ATS. All other lenses including the commercially available in-use control perform statistically similar to the uncoated material control lens. We also observe an interesting flip in dewetting performance for Ampholyte-1 across PBS and ATS, with Ampholyte-1 coated lenses counter-intuitively resisting dewetting in PBS but not in ATS. The interaction of amphiphilic protein molecules in ATS with Ampholyte-1 may generate destabilizing conjoining pressures that drive the dewetting in ATS. Further investigation is necessary to confirm this rationale.  %  (SerMa and MPC are zwitterionic, SerMa attractive interaction with lysosome/mucin?). 

\subsection{Effect of wetting agent surface concentration and combinations on dewetting}\label{WAconc}
Tuning the surface concentration of a wetting agent is an additional parameter that can be utilized for optimizing wetting behavior. With the top performing Ampholyte-1 and Zwitterion-1 coatings, both the film thickness at dewetting and $t_{50p}$ can be controlled by varying wetting agent surface concentration (Fig.\ref{fig:EffectOfConcMixtures}a and Fig.\ref{fig:EffectOfConcMixtures}b). In PBS and ATS, the film thickness at dewetting varies non-linearly with Ampholyte-1 concentration. The film thickness at dewetting for Zwitterion-1 does not show any dependence with surface concentration in PBS, while we observe a beneficial reduction in the critical thickness at higher surface concentrations in ATS. When considering another metric for Ampholyte-1 lenses, $t_{50p}$, we find it increases with surface concentration in PBS, while no such dependence is observed in ATS. For Zwitterion-1, we observe the best performance at the highest concentration, with $t_{50p}$ varying non-linearly in PBS and linearly in ATS.       

Combinations of surface coatings can also be used to enhance wetting characteristics as seen with mixtures of Ampholyte-1 and Zwitterion-1 at different ratios with biocompatible wetting agent GMA (Fig.\ref{fig:EffectOfConcMixtures}c and d). In ATS, the synergistic effect of GMA and Ampholyte-1 is to lower the film thickness at dewetting, whereas no significant improvement was observed in the other systems. GMA also modulated $t_{50p}$, with a 75:25 Zwitterion-1:GMA mixture resisting dewetting of $50\%$ of the lens surface in both PBS and ATS within the time scale of the experiment - the only tested wetting agent to have this beneficial characteristic.

The aforementioned observations are likely driven by variations in contact lens surface morphology and surface energy. For example, the non-linearity in film thickness at dewetting observed in Ampholyte-1 may be an effect of variations in polymer surface conformation that change as a function of surface concentration \cite{dukes2010conformational, milner1991polymer}. A similar effect may also be responsible for the non-linearity in $t_{50p}$ observed with Zwitterion-1. Further studies employing surface microscopy visualizations are necessary to confirm these rationale and identify optimal surface concentrations of wetting agents.

\subsection{Correlation between dewetting performance and dynamic contact angles}\label{surfEnergy} \label{subsec:CorrelationSurfacEnergy}
\begin{figure}[!th]
\centering
\includegraphics[width=\linewidth]{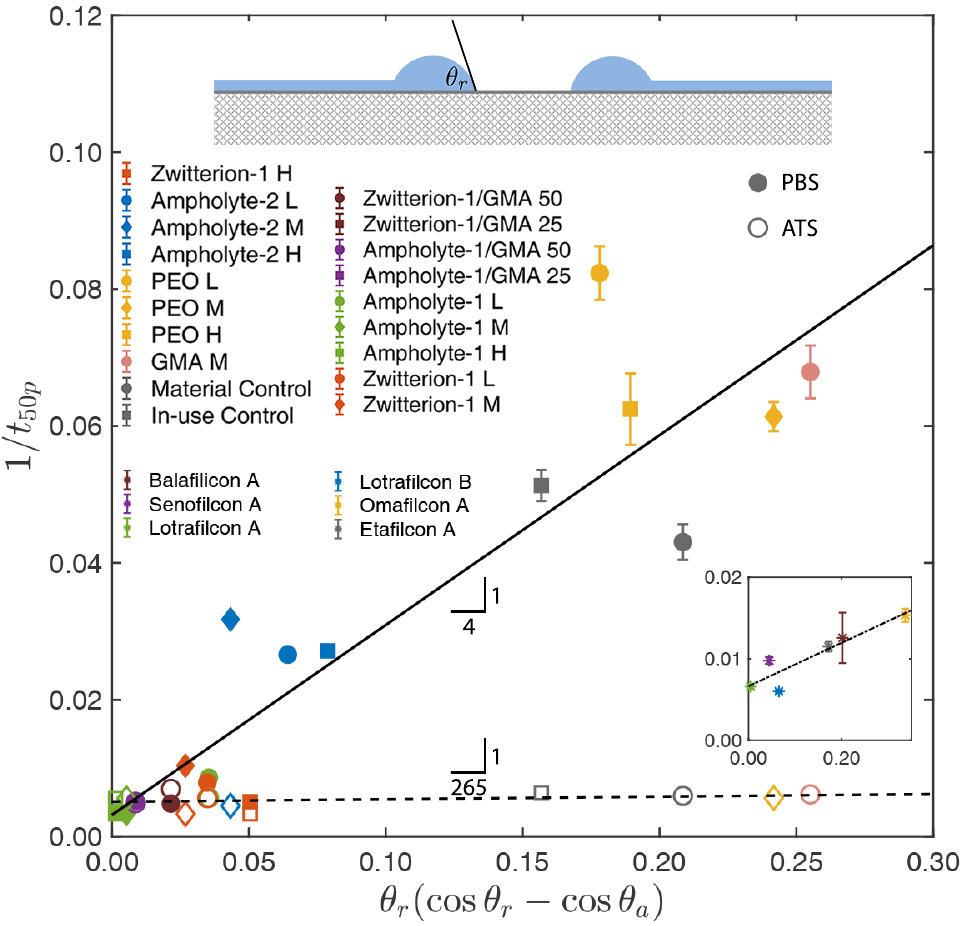}
\caption{$1/t_{50p}$ as a function of the dynamic receding contact angle ($\theta_r$) and the contact angle hysteresis ($\cos \theta_r - \cos \theta_a $). The solid and hollow symbols correspond to PBS and ATS measurements respectively. As predicted by the theory, the experimental data is well described by a linear relationship (goodness of fit for PBS -  $SSE = 0.0019$, $R^2 = 0.8484$). The data presented in the inset correspond to commercial lenses published in literature with contact angle data obtained from Refs. \cite{read2011dynamic,menzies2010vitro} and $t_{50p}$ calculated from data in Ref. \cite{fagehi2013contact}. } \label{fig:ContactAngleTheory} %\textcolor{red}{TO DO: Will ask Kiara to repeat Lysma M}}\label{fig:ContactAngleTheory}
\end{figure}

In this section, we explore quantitative relationships between the contact lens dewetting performance and the contact lens surface energy. We quantify the surface energy through static and dynamic contact angles measured at the three phase contact line (Table \ref{tab:Materials}). The equilibrium (static) contact angle, $\theta_e$, is inversely related to surface energy of the solid \cite{erbil2014debate}. Similarly, the dynamic contact angles (advancing contact angle, $\theta_a$, and receding contact angle, $\theta_r$) are also related to the surface energy; however, they are additionally influenced by the surface roughness. The velocity, $V$, of the dewetted front is a function of the contact angle with $V$ scaling as $\theta_e^3$ \cite{redon1991dynamics}. For non-ideal surfaces with roughness, this relation is known to be modified as $V\propto \theta_r^3$ \cite{de2013capillarity}. Both the above scaling laws are not predictive of the measured $t_{50p}$ (see \bit{Supplementary Materials Fig. S2}).

We obtain a suitable velocity scaling by a force balance between the viscous forces and the uncompensated Young's force at the three phase contact point on the dewetting front \cite{de2013capillarity}:
\begin{equation}\label{eq:ForceBalance}
    \frac{3\mu V \ln_A}{\theta_r}  = \gamma (\cos \theta_r - \cos \bar{\theta}).
\end{equation}
$\mu$ is the dynamic viscosity, $\gamma$ is the surface tension, $\cos \bar{\theta} =\frac{1}{2}(\cos \theta_a + \cos \theta_r)$ and $\ln_A$ accounts for the divergence of viscous dissipation associated with corner flows. Instead of using $\theta_e$ to compute the unbalanced Young's force, we make use of $\bar{\theta}$, which is known to better predict the spreading coefficient of rough surfaces \cite{andrieu1994average}. Expressing $\bar{\theta}$ in terms of dynamic contact angles and simplifying, we obtain the following expression
\begin{equation}\label{eq:charVel}
    V = V^{*} \theta_r (\cos \theta_r - \cos \theta_a),
\end{equation}
where $V^*$ is a characteristic velocity defined as $V^* = \frac{1}{6\ln_A}\frac{\gamma}{\mu}$. 

We compare the predictions from equation \ref{eq:charVel} to our experimental data in Fig.\ref{fig:ContactAngleTheory} by plotting $1/t_{50p}$ (which is proportional to $V$) as a function $\theta_r (\cos \theta_r - \cos \theta_a)$. We obtain sufficiently good agreement between the theory and experiments, with $\frac{1}{t_{50p}}$ varying linearly with $\theta_r (\cos \theta_r - \cos \theta_a)$ as predicted in both PBS and ATS. Similarly, as seen in the inset, the proposed scaling also describe well the data on dewetting of blister pack liquid films over commercial lenses previously published in the literature \cite{read2011dynamic,menzies2010vitro,fagehi2013contact}. We find $V^* \sim 0.3$ (the slope of the best fit line) in the PBS experiments is the same order of magnitude as the predicted $V^* \sim 0.4$ (with $\gamma \sim 0.05\;N/m$, $\mu \sim 0.001$, $\ln_A \sim 20$). Minor discrepancies between the theory and experiments are expected, as we utilize the time since the start of the experiment (Fig.\ref{fig:EffectOfConcMixtures}) instead of the time from onset of dewetting, due to the practical value of the former quantity for contact lens manufacturers. Deviations are expected at low values of $1/t_{50p}$, as additional effects such as evaporation will play a role in film thinning and breakup. In ATS, the measured value ($V^* \sim 0.003$) indicates that capillary forces constitute only a minor role in the dewetting dynamics, with evaporation comprising the dominant influence. Nevertheless, the best performing lenses characterized by large values of $t_{50p}$ also have low values of  $\theta_r (\cos \theta_r - \cos \theta_a)$ in both PBS and ATS.

 $\theta_r (\cos \theta_r - \cos \theta_a)$ is thus a simple metric contact lens manufacturers may use to evaluate the wetting performance. An interesting and impactful corollary is that in addition to the absolute values of the contact angle, the contact angle hysteresis should also minimized to obtain good wetting performance. %Finally, it is also worth noting that surface wetting alteration can compensate for lack of interfacial rheology that accompanies certain opthalmic disorders (such as meibomeium gland dysfunction [cite]) and promote healthy tear film over a blink cycle (based on PBS results)

\section{Conclusions}\label{conc}
In summary, utilizing a previously developed {\it in vitro} drainage and dewetting platform \cite{bhamla2016instability,bhamla2016ddrop}, we established a systematic and improved methodology for identifying optimal wetting agents , and proposed a simple metric to select suitable wetting agents formulations for contact lenses. Notably, the presented study improves on previous works \cite{fagehi2013contact,campbell2013applicability,maldonado2007vitro}  by systematically varying wetting agents on a control base lens, and performing drainage and dewetting experiments that simultaneously characterize terminal drainage,  film thickness at breakup, and dewetting dynamics. Through systematic drainage and dewetting analyses of five wetting agent chemistries across seventeen different formulations applied on the same base lens we revealed that zwitterionic Ampholyte-1 and Zwitterion-1 are effective in reducing film thickness at dewetting and in delaying the propagation of dewetted regions. We also established that mixing of wetting agents and tuning their surface concentration can enhance contact lens wetting characteristics. In particular, a 75:25 Zwitterion-1:GMA wetting agent mixture was observed to resist dewetting in both PBS and ATS over 300 seconds. 
 
%optimal surface coatings and related surface concentrations that can prevent the untimely dewetting of the tear film over a contact lens.
%Valuable addition to existing sessile drop and captive bubble methodology and provides a facile method to simultaneously characteristic both terminal drainage, breakup and dewettting characteristics. 
%In alignment with prior study - ATS delays film drainage.
%Nucleation and growth mechanism
%Finally, through this work, we also establish a reproducible methodology and a simple metric to identify suitable wetting agents for contact lenses.
Extending seminal findings in film dewetting science \cite{redon1991dynamics,andrieu1994average} to clinically relevant measurables, we show that dewetting characteristics across wetting agents with large variations in molecular properties including charge and polarity are well predicted by the dynamic contact angle variations. In particular, we established that the time to dewet $50\%$ of the contact lens linearly scales with the product of the receding contact angle and contact angle hysteresis. This led to the important finding that the combined effect of absolute contact angle value and contact angle hysteresis, which is often overlooked in contact lens dewetting research\cite{fagehi2013contact,campbell2013applicability,maldonado2007vitro,read2011dynamic}, should also be considered during contact lens development for effective contact lens wetting performance. This practical metric is expected to aid academic and industrial researchers in rapidly identifying optimal wetting agent formulations and are expected to better rationalize previous findings \cite{fagehi2013contact,read2011dynamic}. 

Notwithstanding the importance of minimizing tear film dewetting, surface coating solutions must consider a variety of parameters including oxygen permeability, water content, direct patient comfort, and more. While this study primarily focused on dewetting properties, we look forward to further studies which consider additional performance metrics of the tested surface coatings. An important extension of the presented work will be to compare the presented {\it in vitro} data and metrics against {\it in vivo} performance, as the model tear films presented herein approximate, but do not replicate, \textit{in vivo} conditions. Through this more comprehensive approach, contact lenses can be systematically and quickly improved towards optimal molecular and rheological properties, thereby maximizing clinical comfort.

\section*{Appendix}
\subsection{Relative importance of capillary pressure and evaporation in film drainage}

The low Bond number model tear films explored in this manuscript drain under the combined influence of capillary forces and evaporation. The relative importance of the two mechanisms in film drainage can be identified by as follows. 

The film drainage rate under the sole influence of a constant capillary pressure scales as \cite{suja2020foam,bhamla2015influence,mackay1963gravity} 
\begin{equation}\label{eq:capillaryPressureDrainage}
    \frac{dh}{dt} = \frac{d}{dt} \left(\frac{h_0}{\sqrt{1 + \alpha t}}\right) = -\frac{1}{2}\frac{h_0 \alpha}{(1 + \alpha t)^\frac{2}{3}}.
\end{equation}
Here $h_0$ is a reference initial film thickness and $\alpha$ is a capillary drainage time constant.

Similarly, the film thickness evolution under the sole influence of evaporation scales as, 
\begin{equation}\label{eq:EvaporationDrainage}
    \frac{dh}{dt} = \frac{d}{dt} \left( h_0 - Et \right) = -E.
\end{equation}
Here $E$ is the evaporation velocity. Note that these calculation ignores any evaporation induced dynamic phenomenon that can influence drainage\cite{suja2018evaporation}.

Equating equations \ref{eq:capillaryPressureDrainage} and \ref{eq:EvaporationDrainage} and simplifying we obtain,

\begin{equation}
    t_c = \left(\frac{h_0^2}{4\alpha E^2} \right)^\frac{1}{3} - \frac{1}{\alpha} \approx \left(\frac{h_0^2}{4\alpha E^2} \right)^\frac{1}{3}
\end{equation}
Capillary forces dominate film drainage when $t\ll t_c$, whereas evaporation dominates film drainage when $t\gg t_c$. For typical experimentally observed values of $h_0\sim 10\mu m$, $E\sim 0.1\mu m/s$ and $\alpha \sim 1 s^{-1}$, $t_c \sim 30 s$ 

%\section*{References}
\bibliographystyle{vancouver}
\bibliography{References}% Produces the bibliography via BibTeX.
\end{document}